\documentclass[journal]{IEEEtran}
\usepackage{amsmath,amsfonts}
\usepackage{algorithmic}
\usepackage{algorithm}
\usepackage{array}
\usepackage[caption=false,font=normalsize,labelfont=sf,textfont=sf]{subfig}
\usepackage{textcomp}
\usepackage{stfloats}
\usepackage{url}
\usepackage{verbatim}
\usepackage{graphicx}
\usepackage{cite}
\usepackage[utf8]{inputenc}
\usepackage{enumitem} % for custom list settings
\usepackage{titlesec} 
\usepackage{amssymb} 
\usepackage{booktabs} 
\usepackage{threeparttable}

\usepackage{tabularx}
\usepackage{booktabs}
\usepackage{longtable}

\usepackage{listings}
\usepackage{orcidlink}

\usepackage{tikz}
\usepackage{hyperref}
\newcommand{\doublecircle}{%
  \tikz[baseline=-0.5ex]{
    \draw[thick] (0,0) circle [radius=1ex];
    \draw[thick] (0,0) circle [radius=0.65ex];
  }%
}
\definecolor{SkyBlue}{RGB}{135,206,235}        
\definecolor{TealBlue}{RGB}{54,117,136}        
\definecolor{SeaGreen}{RGB}{46,139,87}         
\definecolor{Aquamarine}{RGB}{127,255,212}     
\definecolor{Cerulean}{RGB}{0,123,167}         
\definecolor{CornflowerBlue}{RGB}{100,149,237}
\definecolor{IOT}{RGB}{44,99,168}

\newcommand{\qos}[1]{\textcolor{IOT}{{#1}}}

\newcommand{\appexample}[1]{%
\vspace{0.5em}
\textsc{Application Example (#1):}
}

\begin{document}

% \title{QoS Guard for ROS~2: \\ Automated Validation for Invalid DDS QoS Policy Chains}
\title{Dependency Chain Analysis of ROS~2 DDS QoS Policies: \\From Lifecycle Tutorial to Static Verification}

\author{Sanghoon Lee\,\orcidlink{0000-0002-8160-8952}, Junha Kang\,\orcidlink{0009-0002-4975-0294},  Kyung-Joon Park\,\orcidlink{0000-0003-4807-6461}}
% \thanks{This paper is supported by the Institute of Information and Communications Technology Planning and Evaluation (IITP) Grant funded by Korean Government [Ministry of Science and ICT (MSIT)] under Grant RS-2024-00442085.}
% \thanks{Manuscript received July 9, 2025;}} %revised August 16, 2021.}}

% \markboth{Journal of \LaTeX\ Class Files} %~Vol.~14, No.~8, July~2025}


\maketitle
\begin{abstract}
Robot Operating System~2 (ROS~2) relies on the Data Distribution Service (DDS), which offers more than 20 Quality-of-Service (QoS) policies governing availability, reliability, and resource usage. 
Yet ROS~2 users lack clear guidance on safe policy combinations and validation processes prior to deployment, which often leads to trial-and-error tuning and unexpected run-time failures. 
To address these challenges, we analyze DDS Publisher-Subscriber communication over a life cycle divided into Discovery, Data Exchange, and Disassociation, and provide a user-oriented tutorial explaining how 16 QoS policies operate in each phase. 
Building on this analysis, we derive a QoS dependency chain that formalizes inter-policy relationships and classifies 41 dependency-violation rules, capturing constraints that commonly cause communication failures in practice. 
Finally, we introduce QoS Guard, a ROS~2 package that statically validates DDS XML profiles offline, flags conflicts, and enables safe, pre-deployment tuning without establishing a live ROS~2 session.
Together, these contributions give ROS~2 users both conceptual insight and a concrete tool that enables early detection of misconfigurations, improving the reliability and resource efficiency of ROS~2-based robotic systems.
\end{abstract}


\section{Introduction}
\IEEEPARstart{R}obot Operating System~2 (ROS~2)\cite{ROS2} is the de facto standard platform for robotic and cyber-physical systems across both academia and industry\cite{AR, computing}.   
Serving as middleware between high-level robot applications and the operating system's network stack, ROS~2 enables seamless integration of modular robotic applications and supplies a standardized protocol for both intra- and inter-robot communication~\cite{ROS2_1}.
For the core of the communication, ROS~2 adopts the Data Distribution Service (DDS).
DDS features a Topic-based publish/subscribe model and a decentralized communication~architecture. %UDP-based 

A distinguishing characteristic of DDS is its support for more than 20 tunable Quality-of-Service (QoS) policies, allowing users to fine-tune key system attributes such as availability, reliability, and resource utilization~\cite{DDS}. 
However, despite the richness of available QoS configurations, systematic analysis of their interactions and valid combinations remains relatively limited. 
As a result, ROS~2 users often struggle with the complexity and the large set of QoS policy parameters.

Although the Object Management Group (OMG) specification and various DDS vendors provide definitions and basic usage guidelines for QoS policies, these are often hard to interpret for users without extensive knowledge of DDS. %~\cite{IoT, edge}.
In particular, most current DDS implementations issue only limited warnings for QoS mismatches, while many internal conflicts emerge only later at runtime, appearing as execution errors or performance degradation.
As a result, robot developers are frequently forced to rely on trial-and-error tuning, which can lead to data loss or network saturation in distributed deployments ~\cite{partition}, especially during mission-critical scenarios~\cite{autonomous}.
This challenge comes from two critical gaps.

First, DDS QoS policies rarely function in isolation. 
Because of interdependencies among them, a misconfiguration in one policy can undermine the effectiveness of the others.
However, current standard guidelines and official documentation do not systematically explain the interactions among QoS policies, instead offering only fragmented definitions and isolated recommendations. 

Second, there is a lack of static pre-deployment analysis tools for validating QoS policy profiles. 
Because QoS mismatches of DDS Publisher-Subscriber only surface after communication begins, applications with dozens of nodes-where redeploying hardware in the field is highly impractical-are especially vulnerable~\cite{coorp}. 
In addition, environment-dependent problems such as latency and packet loss reveal themselves only during operation~\cite{Park2025, utilizing}.
It allows a single misconfigured node to escalate into a ``network hotspot" that degrades the performance of the entire robotic platform.

This work addresses these gaps in ROS~2-based robotic development. 
The contributions of this study are as follows:
\begin{itemize}
    \item \textbf{Lifecycle-based QoS Tutorial.}
    We analyze DDS communication across a lifecycle divided into discovery, data Exchange, and disassociation phases, and present a tutorial illustrated with mobile robot examples. 
    It shows how 16 key QoS policies operate in each phase, thereby giving practitioners an intuitive guide.

    \item \textbf{QoS Dependency Chain.}  
    We identify the dependency relationships among QoS policies, classify them into three levels: critical, conditional, and incidental. 
    We then present a novel dependency chain map that offers both academic insight and practical guidance.

    \item \textbf{Static Validation Tool: QoS Guard.}  
    Based on the chain analysis, we extract 41 dependency rules as executable logic and introduce QoS Guard, a static validation tool packaged for ROS~2. 
    QoS Guard scans DDS XML profiles offline, flags conflicts, and enables safe, pre-deployment tuning without requiring live communication.
\end{itemize}

% This paper provides ROS~2 users with an intuitive tutorial on complex DDS QoS policies and, through dependency analysis, delivers a framework that proactively prevents QoS configuration errors.
The remainder of this paper is organized as follows. 
Section~\ref{sec02} provides a structured review of related work on ROS~2 DDS and QoS policies. 
Section~\ref{sec03} introduces the DDS Publisher-Subscriber communication lifecycle and presents a tutorial on 16 core QoS policies based on this lifecycle. 
Section~\ref{sec04} proposes the QoS policy chain, which formalizes the dependencies among QoS policies, and QoS Guard, a static validation tool of DDS XML profiles. 
Section~\ref{sec05} concludes the paper with a summary of the contributions and a discussion of future research directions.
 
\section{Related Work}
\label{sec02}
ROS~2, which has become the de facto standard in the robotics community, adopts DDS as its communication middleware, based on the OMG DDS standard\cite{ROS2}. 
DDS provides a data-centric publish-subscribe model for distributed applications, offering several advantages in robotic communication environments.
The core objective of DDS is to satisfy high performance, predictability, and efficient resource usage within application domains. 
To this end, DDS is designed to minimize dynamic resource allocation and enhance the predictability of resource consumption. 
Another feature of DDS is its set of 22 QoS policies, which allow applications to fine-tune communication behavior and tailor it to their specific requirements~\cite{OMG-DDS}.

Although ROS~2 users can fine-tune communication characteristics using the various QoS policies provided by DDS, they often face significant challenges due to the complexity of the configuration and the risk of potential policy conflicts. 
These difficulties stem primarily from two underlying factors. 
First, DDS QoS policies do not operate independently, but instead exhibit complex and sometimes unpredictable interdependencies. 
Although the OMG DDS specification emphasizes that consistency among policies must be maintained, it offers only vague descriptions such as ``closely related" rather than providing concrete rules, aside from a few basic constraints.

The performance of DDS communication is highly dependent on how QoS policies are combined. 
This has been clearly demonstrated by Fernandez et al.~\cite{Fernande}, who analyzed the effects of various QoS profiles applied in ROS~2.
According to their findings, when \qos{RELIABILITY.kind} QoS is set to \qos{reliable}, an insufficient \qos{HISTORY.depth} value fails to provide an adequate retransmission window, resulting in loss of messages. 
% This confirms a strong dependency between the \qos{RELIABILITY} and the \qos{HISTORY} QoS. 
They also verified that packet latency varies significantly depending on the combination of \qos{RELIABILITY}, \qos{HISTORY}, and \qos{DURABL} settings. 
It shows that communication performance cannot be accurately predicted from a single QoS policy alone, but rather is strongly influenced by how multiple policies interact.

Second, the effectiveness of DDS QoS policies is not independent of the lower layers or the operating environment. 
Although DDS QoS operates at the application and transport layers, the overall end-to-end quality is strongly influenced by delay and packet loss at the network layer~\cite{lee2025probabilistic}.
However, jitter and packet loss, particularly in wireless environments, often remain undetected until the system is actually deployed.
Jaiswa et al.\cite{Jaiswal} reported that in Wi-Fi environments, ROS~2 DDS communication suffers significant throughput degradation when background traffic is present, due to increased NIC buffer delays and increased packet loss. 
Similarly, the study by Dey et al.~\cite{Dey} showed that in wireless multi-robot systems, line-of-sight conditions and robot mobility significantly increase packet loss rates and average latency. 
These results indicate that runtime failures in ROS~2 DDS induced by wireless conditions are difficult to anticipate prior~to~deployment.

To address the issue that QoS configuration errors are typically detected only after deployment, Parra et al.~\cite{Parra} proposed verifying the Required-versus-Offered (RxO) compatibility between ROS~2 Datawriters and Datareaders prior to deployment.
However, this approach is limited to the RxO rules defined by the DDS standard and fails to address intra-policy conflicts, inter-policy interactions, or dynamic dependencies that may emerge under changing operating conditions.
To address these limitations, this study makes three contributions.
First, we present a lifecycle-based tutorial of 16 core QoS policies that function on DDS communication.
Second, we define the QoS Policy Chain, capturing intra-policy conflicts, inter-policy interactions, and dynamic dependencies, and classify them into three levels: critical, conditional, and incidental.
Finally, we extract 41 dependency rules as executable logic in QoS Guard, a static validation tool for ROS~2.
This approach enables ROS~2 users to avoid trial-and-error configuration and proactively prevent conflicts among complex QoS policies.

\section{Lifecycle-based QoS Tutorial}
\label{sec03}
In this section, we define the communication lifecycle between DDS Publishers (DataWriters) and Subscribers (DataReaders) into three distinct phases. Building on this lifecycle, we explain how the QoS policies influence the DDS communication process.

\subsection{Lifecycle of DDS Communication}
\begin{figure}[t]
  \centering
  \includegraphics[width=0.9\linewidth]{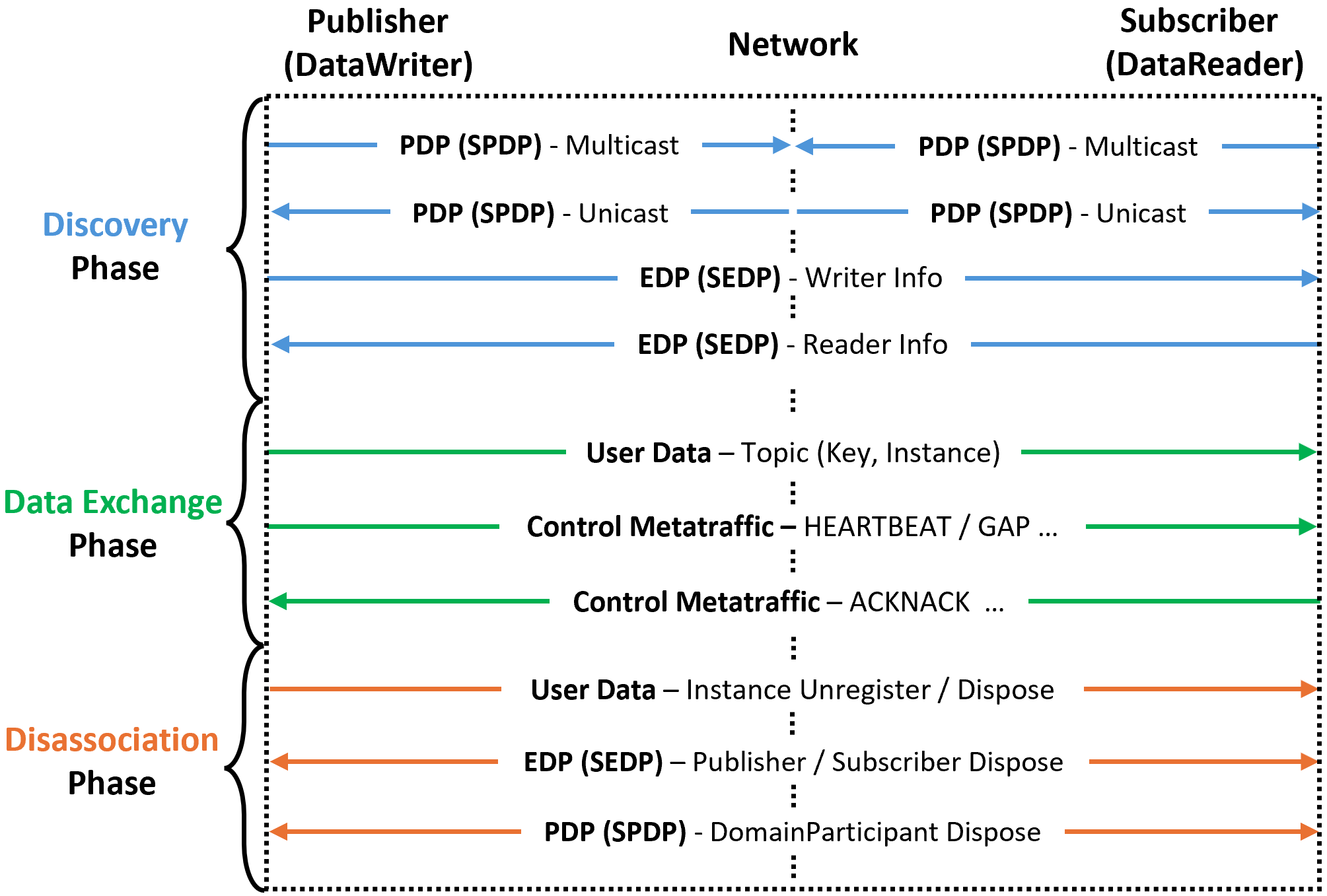}
  \caption{Lifecycle of DDS Communication}
  \label{fig_1}
\end{figure}

DDS communication follows a model in which Publishers and Subscribers exchange data samples under a logical link called a Topic, governed by QoS rules. 
To systematically analyze the impact of QoS in each phase, we divide the communication process between Publishers and Subscribers into three phases: Discovery, Data Exchange, and Disassociation. 
Each of these phases is implemented by DDS vendors based on specific protocols defined in the Real-Time Publish-Subscribe (RTPS) standard, as specified by OMG. 

Figure~\ref{fig_1} illustrates the lifecycle of DDS communication. 
In ROS~2 applications, when a Publisher or Subscriber is created, each first instantiates a DomainParticipant. 
The DomainParticipant serves as a factory for creating DDS entities such as Publishers, Subscribers, and Topics.
Subsequently, the Publisher creates a DataWriter and the Subscriber creates a DataReader for data transmission and reception. 
The Topic binds together a name, a data type, and associated QoS policies, and serves as the logical link between DataWriter and DataReader entities.

\subsubsection{Discovery Phase}
DDS communication begins once these entities have been created.
The first phase of communication is Discovery, during which entities with the same Topic are matched. 
This process is supported in RTPS by its two-layer protocol stack: the Participant Discovery Protocol (PDP) and the Endpoint Discovery Protocol (EDP).
PDP discovers DomainParticipants, while EDP matches DataWriters and DataReaders.

Simple PDP (SPDP), which operates over multicast, is the default in all DDS implementations.
In SPDP, each DomainParticipant periodically multicasts a Participant DATA message, often referred to as an announcement.
A remote participant sends a response with its own Participant DATA message via unicast.
This exchange establishes mutual awareness and network reachability, forming the basis for subsequent endpoint matching.

After SPDP, Simple EDP (SEDP) begins exchanging information about both Topics and endpoints.
Whenever DataWriters, DataReaders are created, modified, or deleted, the DomainParticipant sends SEDP samples to all known participants via unicast. 
The SEDP sample from a DataWriter, known as Writer Info, includes the DomainParticipant and DataWriter GUIDs(Globally Unique Identifiers), offered QoS, Topic details, and a user-data locator (e.g., IP and port).
The DataReader provides a similarly structured message called Reader Info. 
When SEDP samples are received, a QoS compatibility check is performed according to the RxO rules. 
If all conditions are satisfied, a communication pair between the DataWriter and DataReader is established.

\vspace{0.5em}
\subsubsection{Data Exchange Phase}
After the Discovery phase, the DataWriter-DataReader pair begins exchanging user data and control meta-traffic in the Data Exchange phase.
User data, corresponding to the data type defined by a Topic, is transmitted by a DataWriter as serialized payloads encapsulated in DATA or DATA\_FRAG submessages, each identified by a sequence number.
The frequency of user data transmission is entirely determined by the application's write() invocation pattern. 
The DDS middleware, through its QoS policies, is responsible for availability, reliability, ordering, and timeliness.

These service qualities are implemented through control meta-traffic. 
A DataWriter transmits HEARTBEAT and GAP messages: 
HEARTBEAT messages announce the range of sequence numbers currently available, ensuring that DataReaders are aware of which data can be retrieved, while GAP messages indicate sequence numbers that are no longer relevant because they have been filtered or removed. 
Conversely, a DataReader issues ACKNACK and NACK\_FRAG messages to inform the Datawriter of missing samples or fragments, thereby enabling selective retransmission. 
The timing and frequency of these meta-traffic transmissions are defined as configurable parameters in the RTPS standard, and implementations typically expose default values that can be adjusted through profiles or APIs. 
Together, user data and control meta-traffic flow over the same network path, with the latter ensuring availability, reliability, ordering, and timeliness in accordance with communication requirements.
\vspace{1em}

\subsubsection{Disassociation Phase}
During data exchange, if the application stops communication, the DDS may perform disassociation in up to three phases.
The first phase is performed based on the instance, which is the type of data in the Topic.
First, DataWriter can execute a dispose() to indicate that the instance no longer exists, or an unregister() to relinquish responsibility for updating the instance.
These operations are transmitted as a DATA (or DATA\_FRAG) message with specific flags set. 
Upon receiving a sample with the Dispose flag, the DataReader recognizes the instance as inactive and removes the corresponding samples from its cache.
Upon receiving a sample with the Unregister flag, the DataReader recognizes that the DataWriter has relinquished~ownership.

In the second phase, when entities such as Publishers, Subscribers, or Topics are no longer in use, their GUIDs are removed from the matching tables of remote DomainParticipants via SEDP. 
As a result, DataReaders no longer expect further traffic from those endpoints.
In the third phase, when the application terminates its DomainParticipant, periodic Participant DATA announcements sent via SPDP cease. 
If no new announcements are received within a certain timeout period, remote DomainParticipants treat the corresponding GUID as departed and purge all associated history and reliability states.
After these three phases are completed, no GUIDs or samples related to the DomainParticipant remain. 
If the same application restarts a new communication, it starts from the Discovery phase with a newly assigned GUID.

\subsection{QoS Policy in DDS lifecycle} 
This section examines 16 out of the 22 QoS policies defined in the OMG DDS standard, focusing on QoS policies that are practically implemented by the main open-source DDS vendors.
Table~\ref{table1} summarizes the implementation status of each QoS policy in the latest versions of prominent open-source DDS vendors, and indicates which phases of the communication lifecycle each policy affects. 
We then explain how each policy functions in the lifecycle and provide illustrative examples based on a ROS~2-based multi-robot system.
\begin{table*}[htbp]
    \centering
    \begin{threeparttable}
    \caption{Implementation Status and Lifecycle Mapping of 16 DDS QoS Policies}
    \label{table1}
    \renewcommand{\arraystretch}{1.4}
    \begin{tabular}{clcccccccc}
        \toprule
        \# & QoS Policy & Abbr. & \multicolumn{4}{c}{\textbf{Implementation Status}} & \multicolumn{3}{c}{\textbf{Lifecycle Mapping}} \\
        \cmidrule(lr){4-7} \cmidrule(lr){8-10}
           &             &       & ROS~2 & Fast & Cyclone & Open & Discovery & Data Exchange & Disassociation \\
        \midrule
        1  & ENTITY\_FACTORY       & ENTFAC  & $\bigcirc$ & $\bigcirc$ & $\bigcirc$ & $\bigcirc$ & \checkmark &            &             \\
        2  & PARTITION              & PART   & $\bigcirc$ & $\bigcirc$ & $\bigcirc$ & $\bigcirc$ & \checkmark &            &             \\
        3  & USER\_DATA             & USRDATA & $\bigcirc$ & $\bigcirc$ & $\bigcirc$ & $\bigcirc$ & \checkmark &            &             \\
        4  & GROUP\_DATA             & GRPDATA & $\bigcirc$ & $\bigcirc$ & $\bigcirc$ & $\bigcirc$ & \checkmark &            &             \\
        5  & TOPIC\_DATA          & TOPDATA & $\bigcirc$ & $\bigcirc$ & $\bigcirc$ & $\bigcirc$ & \checkmark &            &             \\
        6  & RELIABILITY             & RELIAB  & $\bigcirc$ & $\bigcirc$ & $\bigcirc$ & $\bigcirc$ & \checkmark & \checkmark & \checkmark  \\
        7  & DURABILITY            &  DURABL  & $\triangle$& $\triangle$& $\triangle$& $\bigcirc$ & \checkmark & \checkmark & \checkmark  \\
        8  & DEADLINE                & DEADLN  & $\times$   & $\bigcirc$ & $\bigcirc$ & $\bigcirc$ & \checkmark & \checkmark &             \\
        9  & LIVELINESS              & LIVENS  & $\bigcirc$ & $\bigcirc$ & $\bigcirc$ & $\bigcirc$ & \checkmark & \checkmark & \checkmark  \\
        10 & HISTORY                 &  HIST    & $\bigcirc$ & $\bigcirc$ & $\bigcirc$ & $\bigcirc$ &            & \checkmark &             \\
        11 & RESOURCE\_LIMITS        & RESLIM  & $\bigcirc$ & $\bigcirc$ & $\bigcirc$ & $\bigcirc$ &            & \checkmark &             \\
        12 & LIFESPAN                & LFSPAN  & $\bigcirc$ & $\bigcirc$ & $\bigcirc$ & $\bigcirc$ &            & \checkmark &             \\
        13 & OWNERSHIP (+STRENGTH)   & OWNST   & $\times$   & $\bigcirc$ & $\bigcirc$ & $\triangle$& \checkmark & \checkmark & \checkmark  \\
        14 & DESTINATION\_ORDER      & DESTORD & $\times$   & $\times$   & $\bigcirc$ & $\bigcirc$ & \checkmark & \checkmark &             \\
        15 & WRITER\_DATA\_LIFECYCLE & WDLIFE  & $\times$   & $\times$   & $\bigcirc$ & $\bigcirc$ &            &            & \checkmark  \\
        16 & READER\_DATA\_LIFECYCLE & RDLIFE  & $\times$   & $\times$   & $\times$   & $\bigcirc$ &            &            & \checkmark  \\
        \bottomrule
    \end{tabular}
    \begin{tablenotes}
        \footnotesize
        \item $\bigcirc$ Fully implemented \quad $\triangle$ Partially implemented \quad $\times$ Not implemented
    \end{tablenotes}
    \end{threeparttable}
\end{table*}

\vspace{0.5em}
\subsubsection{ENTITY\_FACTORY}
The \qos{ENTITY\_FACTORY} QoS policy controls how a parent DDS entity-such as a DomainParticipant, Publisher, or Subscriber-creates and initializes its child entities. 
A DomainParticipant has Publishers, Subscribers, and Topics as child entities, while a Publisher has DataWriters and a Subscriber has DataReaders.
This policy has a single boolean parameter, \qos{autoenable\_created\_entities}, which is set to TRUE by default. 
If the value is TRUE, newly created child entities are immediately enabled and begin participating in discovery. 
Conversely, if set to FALSE, the application must explicitly call \qos{enable()} before the entity can participate in discovery.
Thus, the \qos{ENTITY\_FACTORY} QoS serves as a switch that determines when the Discovery phase begins. 
The parameter can be modified at runtime, but changes affect only entities created after the update, not already instantiated.

\appexample{Mobile Robot System} \\
The \qos{ENTITY\_FACTORY} QoS can be used to conserve resources and allow multiple robots to initiate discovery simultaneously under synchronized conditions. 
For instance, the DataWriters and DataReaders of a navigation module may be activated only after completing local sensor calibration or localization. 
By setting \qos{autoenable\_created\_entities=false}, the system delays communication until the robot is ready by explicitly calling \qos{enable()} at the appropriate time. 

\vspace{0.5em}
\subsubsection{PARTITION}
The \qos{PARTITION} QoS policy introduces logical segmentation within a single DDS domain, allowing selective separation or grouping of data flows between Publishers and Subscribers. 
This policy applies to both Publisher and Subscriber entities and has a single parameter: an array of strings, \qos{names}. 
By default, the array contains an empty string.
Partition information is exchanged during the SEDP stage of the Discovery phase. 
A DataWriter and DataReader are matched only if they share at least one common partition name in their respective lists. 
This policy is modifiable at runtime; when the value changes, DDS immediately rematches entities, breaking existing connections and establishing new ones through the SEDP matching procedure.

\appexample{Mobile Robot System} \\
The \qos{PARTITION} QoS can be used to separate identical data types into multiple logical groups without the need to define additional topics or create new domains. 
For instance, delivery robots and inventory robots may share common topics such as ``status'' and ``command'' within the same domain, but still require distinct data flows. 
By setting the \qos{names=delivery} for the delivery robots and \qos{names=inventory} for the inventory robots, a central management system can subscribe only to the desired partition and selectively receive data from a specific group of robots.

\vspace{0.5em}
\subsubsection{USER\_DATA}
The \qos{USER\_DATA} QoS policy allows application-specific metadata to be attached to DDS entities such as DomainParticipant, DataWriter, and DataReader. 
Its parameter is \qos{value}, an arbitrary byte sequence, with the default being an empty sequence. 
DDS does not interpret this sequence or use it for matching; it is simply carried in built-in topic samples (i.e., discovery messages). 
During the Discovery phase, \qos{USER\_DATA} for a DomainParticipant is transmitted in SPDP samples, while that for DataWriters and DataReaders is included in SEDP samples. 
Remote entities can read these samples and use the data in application-level logic. 
This policy can also be updated at runtime, and changes are automatically propagated to remote participants in the next~built-in~topic~sample.

\appexample{Mobile Robot System} \\
The \qos{USER\_DATA} QoS can be used to flexibly deliver identity, authentication, and configuration information without requiring additional topics or separate domains.
For example, each robot may embed \qos{value} such as \qos{robot\_id=R12} and \qos{token=ABCD123} in its participant, allowing the server to inspect the token during the SPDP phase and admit only authorized robots while blocking others. 
Similarly, a DataWriter for a LiDAR topic may include \qos{value} such as \qos{sensor=LiDAR} and \qos{fov=270} in its \qos{USER\_DATA}, enabling the subscribing application to determine sensor configuration and immediately select an appropriate filtering strategy before receiving~any~samples.

\vspace{0.5em}
\subsubsection{GROUP\_DATA}
The \qos{GROUP\_DATA} QoS policy attaches application-specific metadata to the Publisher and Subscriber entities. 
Although it targets different entities, its structure and behavior are identical to those of \qos{USER\_DATA}. 
The \qos{value} of \qos{GROUP\_DATA} is transmitted during the SEDP phase and can also be freely modified at runtime.

\appexample{Mobile Robot System} \\
The \qos{GROUP\_DATA} QoS can be used to logically segment data flows in a manner similar to the \qos{PARTITION} QoS. 
For example, if delivery and inventory robots share topics within the same domain, assigning \qos{value=delivery} or \qos{value=inventory} to each Publisher or Subscriber allows the central management server to read these values during discovery callbacks. 
Although similar to \qos{PARTITION}, the key distinction is that \qos{PARTITION} enforces matching at the DDS level, whereas \qos{GROUP\_DATA} leaves the interpretation of the field entirely to the application.

\vspace{0.5em}
\subsubsection{TOPIC\_DATA}
The \qos{TOPIC\_DATA} QoS policy attaches application-specific metadata to the Topic entity. 
Its format and behavior are identical to those of \qos{USER\_DATA} and \qos{GROUP\_DATA}, and it is propagated during the SEDP phase. 
DDS does not use this field for RxO compatibility matching; rather, it serves as an auxiliary channel for conveying application-specific information. 
This policy can also be freely modified at runtime.

\appexample{Mobile Robot System} \\
The \qos{TOPIC\_DATA} QoS can be used by applications to verify schema compatibility in advance. 
For example, each robot can embed \qos{value} such as \qos{schema=2.1} and \qos{frame=lidar} in the \qos{TOPIC\_DATA} of the \qos{scan\_cloud} topic. 
During topic discovery, an inventory management application can read this information, and if the schema is incompatible, it can prevent subscription and avoid data parsing errors.

\vspace{0.5em}
By combining \qos{USER\_DATA} (Entity), \qos{GROUP\_DATA} (Publisher/Subscriber), and \qos{TOPIC\_DATA} (Topic), a system can hierarchically represent "who (USER), in which group (GROUP), and on which topic (TOPIC) data is being transmitted or received.``
This layered structure provides a flexible and effective mechanism for managing discovery, security, and monitoring in complex robotic systems.

\vspace{0.5em}
\subsubsection{RELIABILITY}
The \qos{RELIABILITY} QoS policy determines whether entities such as Topic, DataWriter, and DataReader transmit data reliably. 
It has two parameters: \qos{kind} and~\qos{max\_blocking\_time}. 
The \qos{kind} parameter is immutable and can be set to either \qos{best\_effort} or \qos{reliable}. 
By default, DataWriter uses \qos{reliable}, while DataReader and Topic use \qos{best\_effort}. 
In \qos{best\_effort} mode, the DataWriter does not wait for ACKs and does not retransmit lost samples. 
In contrast, \qos{reliable} mode attempts to deliver all samples stored in the DataWriter historyCache to the DataReader. 
Retransmissions occur upon request using ACK/NACK signaling, and samples are presented in sequence, meaning that earlier samples must be received before later ones are exposed. 
The \qos{max\_blocking\_time} parameter, used only in \qos{reliable} mode, limits how long a DataWriter's write() or dispose() operation can be blocked due to delayed ACKs or buffer unavailability. 
% Its default value is typically 100 milliseconds.

The \qos{RELIABILITY} QoS policy influences all phases of the DDS communication lifecycle. 
During Discovery, when exchanging SEDP samples, DDS compares the \qos{kind} values of DataWriters and DataReaders according to RxO rules. 
Matching succeeds only if the DataWriter's \qos{kind} is greater than or equal to that of the DataReader 
(\qos{best\_effort}\textless \qos{reliable}).
In the Data Exchange phase, transmission behavior depends on the configured \qos{kind}: in \qos{reliable} mode, the DataWriter periodically sends HEARTBEAT messages, and the DataReader requests missing samples via ACKNACK or NACK\_FRAG, enabling retransmission. 
In \qos{best\_effort} mode, data is transmitted as quickly as possible without additional control mechanisms. 
In the Disassociation phase, dispose and unregister samples are also subject to the same reliability rules.
% A \qos{reliable} DataWriter may fail to complete these operations if they exceed the \qos{max\_blocking\_time}, although this typically does not affect the disassociation logic itself.

\appexample{Mobile Robot System} \\
The \qos{RELIABILITY} QoS can be used to balance safety and efficiency by configuring topics according to their criticality.
For commands such as emergency stops or task assignments, both DataWriter and DataReader should use \qos{reliable}, ensuring guaranteed delivery. 
In contrast, high-frequency data such as LiDAR scans or camera streams can use \qos{best\_effort}, which avoids retransmission overhead and tolerates occasional loss. 

\vspace{0.5em}
\subsubsection{DURABILITY}
The \qos{DURABILITY} QoS policy determines how a late-joining DataReader can receive previously published samples from a DataWriter. 
It has a single parameter of \qos{kind} that supports four possible values. 
The default value, \qos{volatile}, does not send any previous samples to newly joined DataReaders. % even if the DataWriter is still active. 
\qos{Transient\_local} retains samples in the DataWriter's HistoryCache while it is active, allowing late-joining DataReaders to access previously published data. 
Both \qos{transient} and \qos{persistent} preserve data after a DataWriter has been terminated, but the key difference is that \qos{transient} retains data in volatile memory, whereas \qos{persistent} uses non-volatile storage such as files or databases.
The \qos{kind} parameter is immutable and cannot be modified after the entity has been enabled.

The \qos{DURABILITY} QoS policy affects all phases of the DDS communication lifecycle. 
During Discovery, the RxO matching succeeds only if the DataWriter \qos{kind} is greater than or equal to that requested by the DataReader, following the order: \qos{volatile}~$<$~\qos{transient\_local}~$<$~\qos{transient} ~$<$~\qos{persistent}. 
In the Data Exchange phase, a non-volatile DataWriter retransmits retained samples from its HistoryCache to support late-joining DataReaders. 
In the Disassociation phase, if a DataWriter with \qos{transient} or \qos{persistent} durability is deleted or disposes/unregisters an instance, its samples and state are retained by the durability service for future DataReaders.

\appexample{Mobile Robot System}\\
The \qos{DURABILITY} QoS can be used to allow newly added or recovered robots to immediately access the same information, thereby improving both system robustness and collaborative efficiency.
Data such as global maps or mission plans-whose availability must not depend on the lifespan of a specific robot-should remain accessible to robots that join after the publisher has terminated or rebooted. 
In such cases, \qos{DURABILITY} should be set to \qos{transient} if data must persist throughout process restarts, or to \qos{persistent} if it must survive~system-wide~reboots. 

\vspace{0.5em}
\subsubsection{DEADLINE}
The \qos{DEADLINE} QoS policy specifies the maximum interval (\qos{period}, default: infinite) within which a new sample for a given data instance must be produced by the DataWriter and received by the DataReader, as interpreted in the context of their associated Topic.
If the interval is exceeded, DDS raises alarms on both the DataWriter and DataReader to allow the application to detect the deadline miss and react accordingly. 
Thus, \qos{DEADLINE} provides a necessary temporal constraint in scenarios that require regular data updates, such as periodic sensor measurements.

The \qos{DEADLINE} QoS policy affects both the Discovery and Data Exchange phases of the DDS communication lifecycle. 
During Discovery, the RxO matching succeeds only if the DataWriter's \qos{period} is less than or equal to that required by the DataReader.
In the Data Exchange phase, DDS monitors the \qos{period} on a per-instance basis: if the DataWriter fails to meet its deadline, a notification is raised, and if the DataReader does not receive data within the expected interval, a corresponding notification is generated.
These notifications enable the system to detect data delays in real time and execute appropriate recovery or alert mechanisms.

\appexample{Mobile Robot System}\\
The \qos{DEADLINE} QoS can be used for real-time monitoring of robot status. 
Each robot may publish its position and battery level through ROS~2 topics every second. 
The DataWriter and the central monitoring system's DataReader are both configured with a \qos{period} of 1 second.
If a new sample fails to arrive within this interval, the monitoring system receives a deadline-miss notification, enabling immediate detection of communication failures or faults in that data stream. 
The application can then respond by issuing alerts or stopping the robot, thereby enhancing overall system safety and reliability.

\vspace{0.5em}
\subsubsection{LIVELINESS}
The \qos{LIVELINESS} QoS policy enables a DataReader to determine whether its corresponding DataWriter is still active. 
It applies to the Topic, DataWriter, and DataReader entities and defines two parameters: \qos{kind} and \qos{lease\_duration}.
The \qos{kind}, specifies who is responsible for asserting liveliness. 
In the default mode, \qos{automatic}, the DomainParticipant asserts liveliness implicitly by periodical liveliness assertions.
The \qos{manual\_by\_participant} mode, a single assertion from any entity within a DomainParticipant marks all of its DataWriters as alive. 
In \qos{manual\_by\_topic} mode, each DataWriter must explicitly assert its own liveliness by publishing HEARTBEAT samples or calling assert\_liveliness(). 
The second parameter, \qos{lease\_duration}, defines the maximum time a DataReader waits after missing liveliness assertions before declaring the DataWriter as not alive; its default is infinite.

The \qos{LIVELINESS} QoS policy influences all phases of the DDS communication lifecycle. 
During Discovery, the RxO matching succeeds only if the DataWriter's \qos{kind} is greater than or equal to that required by the DataReader (\qos{automatic}$<$\qos{manual\_by\_participant}$<$\qos{manual\_by\_topic}), and its \qos{lease\_duration} is shorter than or equal to the DataReader's value.
In the Data Exchange phase, the DataWriter sends periodic liveliness assertions or explicitly calls assert\_liveliness(). 
In \qos{manual\_by\_topic} mode, the liveliness is indicated by setting the LivelinessFlag in HEARTBEAT message. 
In the Disassociation phase, if no liveliness assertion is received within the configured \qos{lease\_duration}, the DataReader marks the DataWriter as not alive. 
If all writers disappear, the instance state transitions to \qos{not\_alive\_no\_writers}.

\appexample{Mobile Robot System}\\
The \qos{LIVELINESS} QoS can be used to verify whether the publishing process itself is still active, whereas the \qos{DEADLINE} QoS ensures the timely delivery of individual data samples. 
This policy enables a central monitoring system to automatically track the operational status of each robot.
For instance, each robot may publish position and battery level via a DataWriter. 
The central DataReader is configured with \qos{kind} set to \qos{automatic} and \qos{lease\_duration} set to 5 seconds, causing DDS to refresh liveliness every five seconds.
If a signal is not received within the lease duration, a liveliness notification is triggered to indicate that the robot is inactive. 
The monitoring application can then respond by graying out the robot's icon on the map or issuing a warning. 

\vspace{0.5em}
\subsubsection{HISTORY}
The \qos{HISTORY} QoS policy determines how many samples a DataWriter retains in its HistoryCache for retransmission, and how many samples a DataReader stores before they are delivered to the application. 
This policy applies to Topic, DataWriter, and DataReader entities and must be configured at the entity creation time, as it cannot be modified afterward. 
The \qos{kind} parameter supports two values: \qos{keep\_last}, which retains only the most recent samples per instance, with \qos{depth} specifying the maximum number of samples to keep.
By default, \qos{depth=1}, so only the latest sample is delivered.
\qos{keep\_all} stores all samples for each instance and attempts to deliver as many as possible; in this case, the \qos{depth} parameter is ignored.

The \qos{HISTORY} QoS policy plays a central role during the Data Exchange phase. 
Both DataWriter and DataReader maintain their own HistoryCache, and data exchange essentially becomes the process of synchronizing these caches. 
When a DataWriter calls write(), a new sample (CacheChange) is appended to its HistoryCache and transmitted via RTPS into the DataReader's HistoryCache. 
The configured \qos{kind} and \qos{depth} determine how many samples are retained in each cache, thereby enforcing whether only the latest samples (\qos{keep\_last}) or all samples (\qos{keep\_all}) are preserved.

\appexample{Mobile Robot System}\\
The \qos{HISTORY} QoS can be used to control how much robot data is retained. 
If a control station must preserve all robot positions since startup, \qos{kind=keep\_all} can be set so that both the DataWriter and the DataReader store every sample. 
In contrast, for real-time tracking where only the latest position matters, setting \qos{kind=keep\_last} with \qos{depth=1} ensures that each robot's DataWriter retains and transmits only the most recent position, while the DataReader receives only that single value.

\vspace{0.5em}
\subsubsection{RESOURCE\_LIMITS}
The \qos{RESOURCE\_LIMITS} QoS policy defines upper bounds on the number of instances and samples that the Topic, DataWriter, and DataReader entities can manage. 
An instance is a specific data object within a Topic, identified by its key field, while a sample is a transmission unit that contains both the data of an instance and its associated metadata. 
This policy is immutable after configuration and specifies three parameters. 
The \qos{max\_samples} parameter limits the total number of samples across all instances; 
the \qos{max\_instances} parameter restricts the maximum number of instances; and the \qos{max\_samples\_per\_instance} parameter limits the number of samples per instance. 
Although the OMG standard sets all three parameters to unlimited by default, some DDS implementations impose default limits to protect memory resources.

The \qos{RESOURCE\_LIMITS} QoS policy directly restricts the amount of data a DataWriter or DataReader can store in its HistoryCache during the Data Exchange phase. 
The HistoryCache is a logical buffer associated with each entity. 
While the \qos{HISTORY} QoS defines how samples are preserved, the \qos{RESOURCE\_LIMITS} parameters set bounds on the total number of samples and instances, thereby preventing cache overflow and memory exhaustion.

\appexample{Mobile Robot System}\\
The \qos{RESOURCE\_LIMITS} QoS can be used to manage resources and maintain communication stability. 
For instance, where only the most recent information is important and historical records are less critical, such as a robot's real-time position, the DataWriter's \qos{max\_instances} can be set to a small value to avoid excessive memory usage. 
On the DataReader side, where the number of participating robots may be large or variable, the \qos{max\_samples\_per\_instance} value should be set high enough to keep minimal data for each robot instance.

\vspace{0.5em}
\subsubsection{LIFESPAN}
The \qos{LIFESPAN} QoS policy defines how long a sample published by a DataWriter remains valid. 
It applies to Topic and DataWriter entities, and may also apply to a DataReader depending on the DDS implementation. 
This policy has a single parameter, \qos{duration}, which defaults to infinity. 
When a DataWriter sends a sample, the expiration time is calculated by adding the \qos{duration} to the timestamp. 
Once this time has elapsed, the sample is automatically removed from both the DataWriter's and DataReader's HistoryCache. 
As a mutable QoS policy, \qos{LIFESPAN} can be modified after the entity is enabled, allowing the application to adjust the sample lifetimes based on real-time requirements.

The \qos{LIFESPAN} QoS policy directly affects the Data Exchange phase. 
Each time a DataWriter publishes a sample, its expiration is tracked, and once expired, the sample is automatically removed from the HistoryCache, preventing access by the DataReader. 
This ensures that DataReader always accesses up-to-date information and helps to manage memory usage within the HistoryCache.

\appexample{Mobile Robot System}\\
The \qos{LIFESPAN} QoS can be used to prevent robots from retaining outdated samples unnecessarily. 
For data such as position or battery level, where only the last few seconds matter, setting the \qos{duration} accordingly ensures that the DataWriter's history cache stores only the most recent samples, with older ones automatically deleted to conserve memory. 
In contrast, for data such as command logs -- where long-term delivery is more important than freshness -- it is preferable to set \qos{duration} to infinity and rely on other QoS policies to ensure reliability.

\vspace{0.5em}
\subsubsection{OWNERSHIP (+STRENGTH)}
The \qos{OWNERSHIP} QoS policy determines whether multiple DataWriters can concurrently update the same instance, or, if not, which DataWriter's value should be accepted. 
It applies to Topic, DataWriter, and DataReader entities and is immutable once the entity has been enabled. 
The \qos{kind} parameter has two possible values. 
The default value, \qos{shared}, allows multiple DataWriters to freely update the same instance. 
In contrast, \qos{exclusive} enforces that only a single DataWriter is allowed to update an instance, and its updates alone are delivered to DataReaders.
Ownership can change dynamically, and priority is determined by \qos{value} of the \qos{OWNERSHIP\_STRENGTH} QoS policy. 
This auxiliary policy is meaningful only in \qos{exclusive} mode.
It assigns to each DataWriter an integer strength \qos{value} (default: 0) and is mutable even after activation.

The \qos{OWNERSHIP} and \qos{OWNERSHIP\_STRENGTH} QoS policies influence all phases of the DDS communication lifecycle. 
During Discovery, RxO matching succeeds only if the \qos{kind} values of the DataWriter and DataReader are identical.
In the Data Exchange phase, the configured \qos{kind} determines how instances flow to the DataReader. 
In \qos{shared} mode, the DataReader receives updates from all DataWriters, whereas in \qos{exclusive} mode, only samples from the highest-priority DataWriter are delivered. 
Finally, in the Disassociation phase, if a DataWriter calls dispose() or unregister(), the state change is reflected in the DataReader only if the request comes from the current owner in \qos{exclusive} mode.

\appexample{Mobile Robot System}\\
The \qos{OWNERSHIP} and \qos{OWNERSHIP\_STRENGTH} QoS policies can be used to manage shared resources or mission states consistently. 
For example, if multiple robots scan the environment simultaneously to build a shared map, \qos{shared} mode allows the server to receive updates from all robots and generate a unified map.
In contrast, for tasks that must be performed by a single robot, \qos{exclusive} mode can be used with appropriate \qos{value} assigned to each DataWriter, ensuring that the active robot becomes the instance owner. 
If the current owner robot fails to respond due to a fault, DDS automatically transfers ownership to the standby robot with the next highest \qos{value}, allowing the task to continue without interruption.

\vspace{0.5em}
\subsubsection{DESTINATION\_ORDER}
The \qos{DESTINATION\_ORDER} QoS policy determines how a DataReader orders samples from multiple DataWriters targeting the same instance. 
It applies to Topic, DataWriter, and DataReader entities and is immutable once the entity is enabled. 
The \qos{kind} parameter has two values. 
The default, \qos{by\_reception\_timestamp}, orders samples by their reception time at the DataReader. 
In contrast, \qos{by\_source\_timestamp} uses the timestamp assigned by the DataWriter at publication, preserving the original creation order regardless of delays.

The \qos{DESTINATION\_ORDER} QoS policy affects both the Discovery and Data Exchange phases. 
During Discovery, RxO matching succeeds only if the \qos{kind} offered by the DataWriter is greater than or equal to that requested by the DataReader (\qos{by\_reception\_timestamp}$<$\qos{by\_source\_timestamp}). 
In the Data Exchange phase, the \qos{kind} setting governs the processing order of samples that arrive at the DataReader from multiple DataWriters for the same instance.

\appexample{Mobile Robot System}\\
The \qos{DESTINATION\_ORDER} QoS policy can be used to maintain data consistency when multiple robots simultaneously update the same instance. 
By configuring \qos{by\_source\_timestamp}, samples are ordered by their creation time regardless of network delays or arrival order, allowing all robots to share a consistent view of the map. 
In contrast, for real-time data such as current positions, where the latest received value is most important, using \qos{by\_reception\_timestamp} enables the DataReader to reflect the earliest arriving sample immediately.

\vspace{0.5em}
\subsubsection{WRITER\_DATA\_LIFECYCLE}
The \qos{WRITER\_DATA\_LIFECYCLE} QoS policy defines whether a DataWriter should notify DataReaders with a dispose() when it unregisters an instance. 
It applies only to DataWriter entities and is a mutable policy that can be changed at runtime. 
Its sole parameter, \qos{autodispose\_unregistered\_instances} (default: true), causes an instance to be automatically marked as disposed when an unregister() is called, so that it is recognized as deleted on the DataReader side. 
If set to false, the unregister() only disassociates the writer from the instance without disposing of it; the application must explicitly call dispose() to fully delete the instance.

The \qos{WRITER\_DATA\_LIFECYCLE} QoS policy affects the Disassociation phase of the DDS communication lifecycle. 
When a DataWriter calls unregister() or is deleted, the value of \qos{autodispose\_unregistered\_instances} determines how the DataReader interprets the instance state. 
If set to true, the instance is immediately reported as \qos{not\_alive\_disposed} and treated as fully removed. 
Conversely, if set to false, only the association between the DataWriter and the instance is removed, and the instance is marked as \qos{not\_alive\_no\_writers} if no other writers exist.

\appexample{Mobile Robot System}\\
The \qos{WRITER\_DATA\_LIFECYCLE} QoS can be used to explicitly manage the lifecycle of object-based tasks. 
For example, when a robot detects an object with its sensors, it can publish the object's key, position, type, and status on a topic, allowing other robots or the control system to subscribe and build a shared environment model. 
When the task is completed, the post-processing behavior depends on the detecting robot's \qos{autodispose\_unregistered\_instances} setting. 
If true, the instance is immediately disposed of and removed from the map or marked as ``processed''; if false, only the Datawriter-instance link is removed while the object information remains active, allowing another robot to rediscover and update the same object, thereby improving collaboration flexibility.

\vspace{0.5em}
\subsubsection{READER\_DATA\_LIFECYCLE}
The \qos{READER\_DATA\_LIFECYCLE} QoS policy determines how long a DataReader retains samples that have been disposed of or are no longer associated with any DataWriter. 
This policy applies exclusively to DataReader entities and is mutable, allowing updates at runtime. 
It defines two parameters. 
The first, \qos{autopurge\_disposed\_samples\_delay}, specifies how long samples and metadata are retained after an instance transitions to the \qos{not\_alive\_disposed} state. 
The second, \qos{autopurge\_no\_writer\_samples\_delay}, defines the maximum time to retain samples after an instance enters the \qos{not\_alive\_no\_writers} state, indicating that all associated writers have disappeared. 
Both parameters default to infinity. 
Once the specified time has elapsed, the DataReader automatically purges the instance and its related samples to reclaim memory.

The \qos{READER\_DATA\_LIFECYCLE} QoS policy is relevant during the Disassociation phase of the DDS communication lifecycle. 
It governs when a DataReader should automatically remove samples of inactive instances. 
When a DataWriter calls dispose(), the DataReader marks the instance as \qos{not\_alive\_disposed}; 
after unregister() is called and no other writer remains, the instance is marked as \qos{not\_alive\_no\_writers}. 
Both states indicate that the instance is no longer active. 
Once the corresponding timer--either \qos{autopurge\_disposed\_samples\_delay} or \qos{autopurge\_no\_writer\_samples\_delay}--expires, the DataReader automatically purges the instance. 
This mechanism allows the DataReader to autonomously complete cleanup after a writer-side termination signal, preventing memory leaks and maintaining system integrity.

\appexample{Mobile Robot System}\\
The \qos{READER\_DATA\_LIFECYCLE} QoS policy enables tiered instance management for improved efficiency. 
For example, in a temporary storage area where hundreds of pallets move quickly and historical positions become obsolete immediately, setting \qos{autopurge\_disposed\_samples\_delay} to 0~seconds ensures that the cache is cleared as soon as a robot calls dispose(). 
This keeps the cache lightweight and prevents unnecessary memory growth. 
In contrast, for static, critical objects awaiting inspection after relocation, setting \qos{autopurge\_no\_writer\_samples\_delay} to 300~seconds allows newly joined robots to continue inspection even after brief communication interruptions. 
By adjusting the reader-side delay values based on context, essential data can be preserved while irrelevant information is promptly discarded, ensuring efficient use of system resources.

\section{QoS Policy Chain Analysis \& QoS Guard}
\label{sec04}
Based on the analysis of the operational logic of standard QoS policies across the DDS communication lifecycle in Section~\ref{sec03}, this section identifies the dependency relationships among 16 core QoS policies and proposes a unified structure called the ``QoS Policy Chain.'' 
Building on this chain, we also introduce QoS Guard, an automated tool that validates user-defined XML QoS profiles, detects potential issues from inter-policy dependencies, and suggests corrective actions.

\subsection{QoS Policy Chain Analysis}
We classify QoS dependencies into three categories: Critical Dependency, Conditional Dependency, and Incidental Dependency. 
A Critical Dependency (A $\rightarrow$ B) occurs when QoS policy~A must satisfy a specific configuration before QoS policy~B can operate, or when the combination is explicitly disallowed by the DDS standard or implementation. 
A Conditional Dependency (A $\rightarrow$ B) arises when QoS policy~A's configuration may impair the correct behavior of QoS policy~B-such as causing loss of guarantees, runtime errors, or performance degradation-without preventing B from operating entirely. 
An Incidental Dependency (A $\rightarrow$ B) refers to cases where QoS policy~A indirectly affects the behavior or performance of QoS policy~B without violating its guarantees or functionality.

Table~\ref{table2} summarizes the dependency relationships among the 16 QoS policies. 
For readability, the names of QoS policies in this section are replaced with abbreviations, which are listed in Table~\ref{table1}.
Dependency directions are denoted as forward~($\rightarrow$), reverse~($\leftarrow$), or bidirectional~($\leftrightarrow$), while dependency strength is represented using the following symbols: Critical~($\text{X}$), Conditional~(\doublecircle{}), and Incidental~($\bigcirc$). 
This table facilitates quick identification of policy combinations that may cause critical conflicts or, under certain conditions, lead to quality degradation or performance issues. 
The subsequent paragraphs provide a step-by-step analysis of the specific dependencies associated with each QoS policy.

\renewcommand{\arraystretch}{1.5}

\begin{table*}[htbp]
\centering
\caption{Dependency relationships among 16 QoS policies.}
\label{table2}
\resizebox{\textwidth}{!}{%
\begin{tabular}{|c|c|c|c|c|c|c|c|c|c|c|c|c|c|c|c|c|}
\hline
\textbf{\#} & {ENTFAC} & {PART} & {USRDATA} & {GRPDATA} & {TOPDATA} & {RELIAB} & {DURABL} & {DEADLN} & {LIVENS} & {HIST} & {RESLIM} & {LFSPAN} & {OWNST} & {DESTORD} & {WDLIFE} & {RDLIFE} \\
\hline
{ENTFAC} & -- & -- & -- & -- & -- & -- & $\bigcirc~\rightarrow$ & -- & -- & -- & -- & -- & -- & -- & -- & -- \\
\hline
{PART} & -- & $\text{X}$ \  $\leftrightarrow$ & -- & -- & -- & -- & $\bigcirc~\rightarrow$ & $\bigcirc~\rightarrow$ & $\bigcirc~\rightarrow$ & -- & -- & -- & -- & -- & -- & -- \\
\hline
{USRDATA} & -- & -- & -- & -- & -- & -- & -- & -- & -- & -- & -- & -- & -- & -- & -- & -- \\
\hline
{GRPDATA} & -- & -- & -- & -- & -- & -- & -- & -- & -- & -- & -- & -- & -- & -- & -- & -- \\
\hline
{TOPDATA} & -- & -- & -- & -- & -- & -- & -- & -- & -- & -- & -- & -- & -- & -- & -- & -- \\
\hline
{RELIAB} & -- & -- & -- & -- & -- & $\text{X}$ \  $\leftrightarrow$ & $\text{X}$ \  $\rightarrow$ & $\doublecircle~\rightarrow$ & $\doublecircle~\rightarrow$ & $\doublecircle~\leftarrow$ & $\doublecircle~\leftarrow$ & $\doublecircle~\leftarrow$ & $\text{X}$ \  $\rightarrow$ & -- & $\doublecircle~\rightarrow$ & -- \\
\hline
{DURABL} & $\bigcirc~\leftarrow$ & $\bigcirc~\leftarrow$ & -- & -- & -- & $\text{X}$ \  $\leftarrow$ & $\text{X}$ \  $\leftrightarrow$ & $\bigcirc~\rightarrow$ & -- & $\doublecircle~\leftarrow$ & $\doublecircle~\leftarrow$ & $\doublecircle~\leftarrow$ & -- & -- & -- & $\bigcirc~\leftarrow$ \\
\hline
{DEADLN} & -- & $\bigcirc~\leftarrow$ & -- & -- & -- & $\doublecircle~\leftarrow$ & $\bigcirc~\leftarrow$ & $\text{X}$ \ ~$\leftrightarrow$ & $\doublecircle~\leftarrow$ & -- & -- & -- & $\doublecircle~\rightarrow$ & -- & -- & -- \\
\hline
{LIVENS} & -- & $\bigcirc~\leftarrow$ & -- & -- & -- & $\doublecircle~\leftarrow$ & -- & $\doublecircle~\rightarrow$ & $\text{X}$ \  $\leftrightarrow$ & -- & -- & -- & $\doublecircle~\rightarrow$ & -- & -- & $\doublecircle~\rightarrow$ \\
\hline
{HIST} & -- & -- & -- & -- & -- & $\doublecircle~\rightarrow$ & $\doublecircle~\rightarrow$ & -- & -- & -- & $\text{X}$ \  $\leftrightarrow$ & $\doublecircle~\leftrightarrow$ & -- & $\doublecircle~\leftrightarrow$ & -- & -- \\
\hline
{RESLIM} & -- & -- & -- & -- & -- & $\doublecircle~\rightarrow$ & $\doublecircle~\rightarrow$ & -- & -- & $\text{X}$ \  $\leftrightarrow$ & $\text{X}$ \  $\leftrightarrow$ & $\doublecircle~\leftrightarrow$ & -- & $\doublecircle~\rightarrow$ & -- & -- \\
\hline
{LFSPAN} & -- & -- & -- & -- & -- & $\doublecircle~\rightarrow$ & $\doublecircle~\rightarrow$ & -- & -- & $\doublecircle~\leftrightarrow$ & $\doublecircle~\leftrightarrow$ & -- & -- & -- & -- & -- \\
\hline
{OWNST} & -- & -- & -- & -- & -- & $\text{X}$ \  $\leftarrow$ & -- & $\doublecircle~\leftarrow$ & $\doublecircle~\leftarrow$ & -- & -- & -- & $\text{X}$ \  $\leftrightarrow$ & -- & $\bigcirc~\rightarrow$ & -- \\
\hline
{DESTORD} & -- & -- & -- & -- & -- & -- & -- & -- & -- & $\doublecircle~\leftarrow$ & $\doublecircle~\leftarrow$ & -- & -- & $\text{X}$ \  $\leftrightarrow$ & -- & -- \\
\hline
{WDLIFE} & -- & -- & -- & -- & -- & $\doublecircle~\leftarrow$ & -- & -- & -- & -- & -- & -- & $\bigcirc~\leftarrow$ & -- & -- & $\doublecircle~\rightarrow$ \\
\hline
{RDLIFE} & -- & -- & -- & -- & -- & -- & $\bigcirc~\rightarrow$ & -- & $\doublecircle~\leftarrow$ & -- & -- & -- & -- & -- & $\doublecircle~\leftarrow$ & -- \\
\hline
\end{tabular}%
}
\end{table*}

First, the five QoS policies that operate exclusively during the Discovery phase--\qos{ENTFAC}, \qos{PART}, \qos{USRDATA}, \qos{GRPDATA}, and \qos{TOPDATA}--do not depend on any other QoS policies. 
The \qos{ENTFAC} policy is involved only in the initial entity creation stage of the DDS lifecycle, where no other QoS policies apply.
The \qos{PART} policy has only a Critical self-dependency, in which it checks the compatibility of its \qos{values} during SEDP RxO matching. 
The \qos{USRDATA}, \qos{GRPDATA}, and \qos{TOPDATA} policies are propagated via SEDP samples, but their actual interpretation and use are entirely determined by the ROS~2 application layer. 
Since DDS merely forwards this data without interpreting its semantics, these policies are also completely independent of other QoS policies.

The \qos{RELIAB} and \qos{DURABL} QoS policies ensure topic availability and consistency and are involved in all phases of the DDS communication lifecycle. 
Specifically, the \qos{RELIAB} QoS policy exhibits a Critical self-dependency in SEDP RxO matching and has Conditional dependencies on \qos{HIST}, \qos{RESLIM}, and \qos{LFSPAN}. 
The retransmission mechanism in \qos{reliable} mode relies on the DataWriter's History Cache, whose retention behavior and lifetime are governed by these QoS policies.
If \qos{HIST.kind} is set to \qos{keep\_last} with a \qos{HIST.depth} that is too small, write() operations may block or result in sample drops, thus compromising reliability guarantees. 

In \qos{HIST.kind=keep\_all}, the HistoryCache size is constrained by \qos{RESLIM} parameters such as \qos{max\_samples}, which can cause similar issues. 
Conversely, setting \qos{RESLIM.max\_samples} too high can trigger burst retransmissions of old samples under link outages and high packet-loss conditions, overloading memory and network resources and resulting in substantial delays~\cite{lee2025optimizing}.
If the \qos{LFSPAN.duration} is set to be shorter than the round-trip time (RTT) plus the retransmission window, samples may expire before retransmission, effectively nullifying the guarantees of \qos{RELIAB}. 
Therefore, the values of these three QoS policies must be configured properly to ensure reliable operation when \qos{RELIAB.kind=reliable}.

\begin{figure}[t]
\centering
\includegraphics[width=0.91\linewidth]{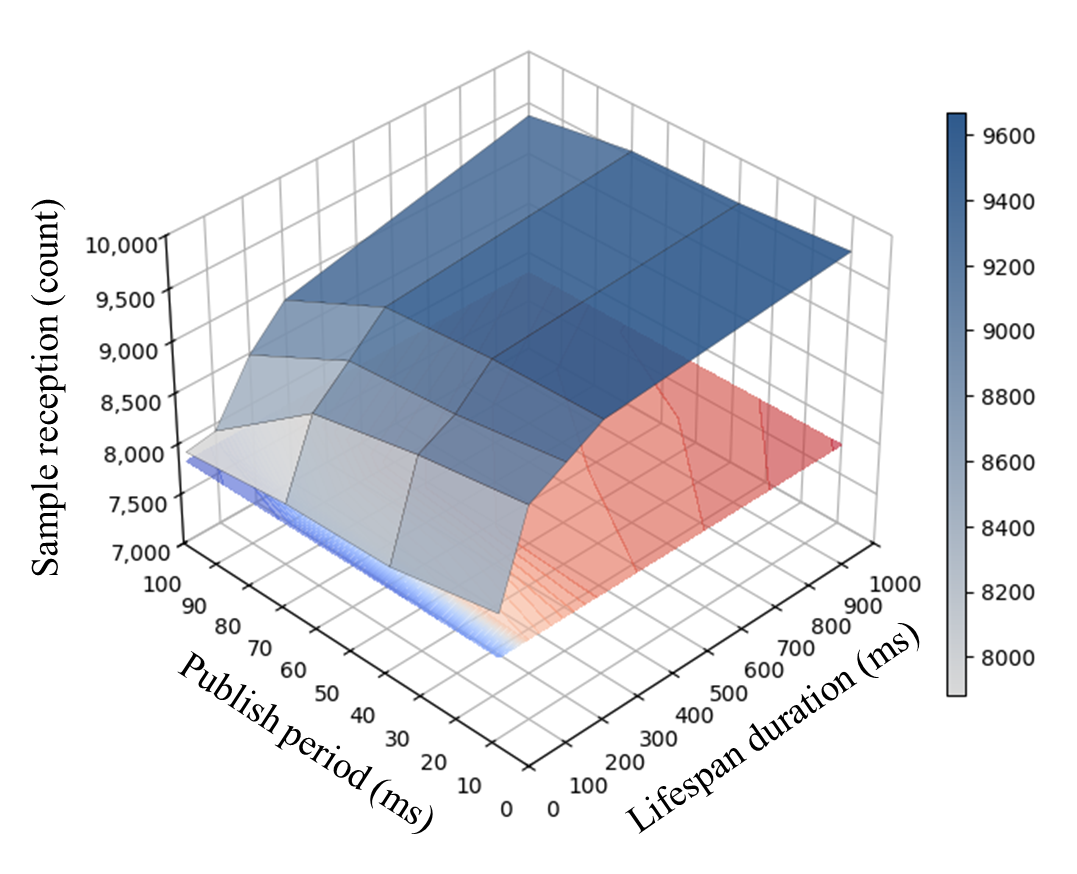}
\caption{Impact of LIFESPAN and Publish Period on Sample Reception}
\label{fig_3}
\end{figure}

Figure~\ref{fig_3} visualizes the impact of \qos{LFSPAN} on reliability. 
The experiment was conducted with a Publisher-Subscriber pair configured with \qos{RELIAB.kind=reliable}, while \qos{HIST.depth} was fixed at its maximum. 
The communication link was subjected to 20\% packet error rate. 
The graph shows how the number of messages transmitted and received varies depending on the \qos{LFSPAN.duration} and the publish period. 
Regardless of the \qos{HIST.depth}, samples are deleted once their \qos{LFSPAN} expires. 
As shown in the dependency analysis, samples may expire before the retransmission window opens, causing \qos{reliable} to behave effectively like \qos{best\_effort}. 
In contrast, if the \qos{LFSPAN.duration} is set sufficiently longer than the publish period, retransmission opportunities are preserved even under packet loss, enabling a high delivery rate. 
This occurs because DDS allows repeated retransmission sessions as long as the \qos{LFSPAN.duration} remains valid.

% In conclusion, the \qos{LFSPAN} configuration must be carefully chosen with respect to both the publish rate and the packet loss rate of the link. If \qos{LFSPAN.duration} is too short, it can improve memory efficiency but significantly reduce reliability; if it is too long, it can unnecessarily consume memory and network resources for old data. Striking the right balance is therefore essential to satisfy both data integrity and resource efficiency.

The \qos{DURABL} QoS policy exhibits Critical dependencies on itself in SEDP RxO matching and on \qos{RELIAB}. 
Major DDS implementations require that when \qos{DURABL.kind$\geq$transient\_local}, the DataWriter must be configured with \qos{RELIAB.kind=reliable}. 
This ensures that late-joining DataReaders can reliably receive previous samples.
The \qos{DURABL} QoS policy also has Conditional dependencies on \qos{HIST}, \qos{RESLIM}, and \qos{LFSPAN}, since it relies on the DataWriter's History Cache for transmission. 
If the History Cache is too small or the \qos{LFSPAN} duration is too short, previous samples may not be retained, nullifying the intended effect of \qos{DURABL}. 
Conversely, if these values are excessively large, they do not violate functionality but may increase memory and network usage, potentially leading to higher~latency.

\begin{figure}[t]
\centering
\includegraphics[width=\linewidth]{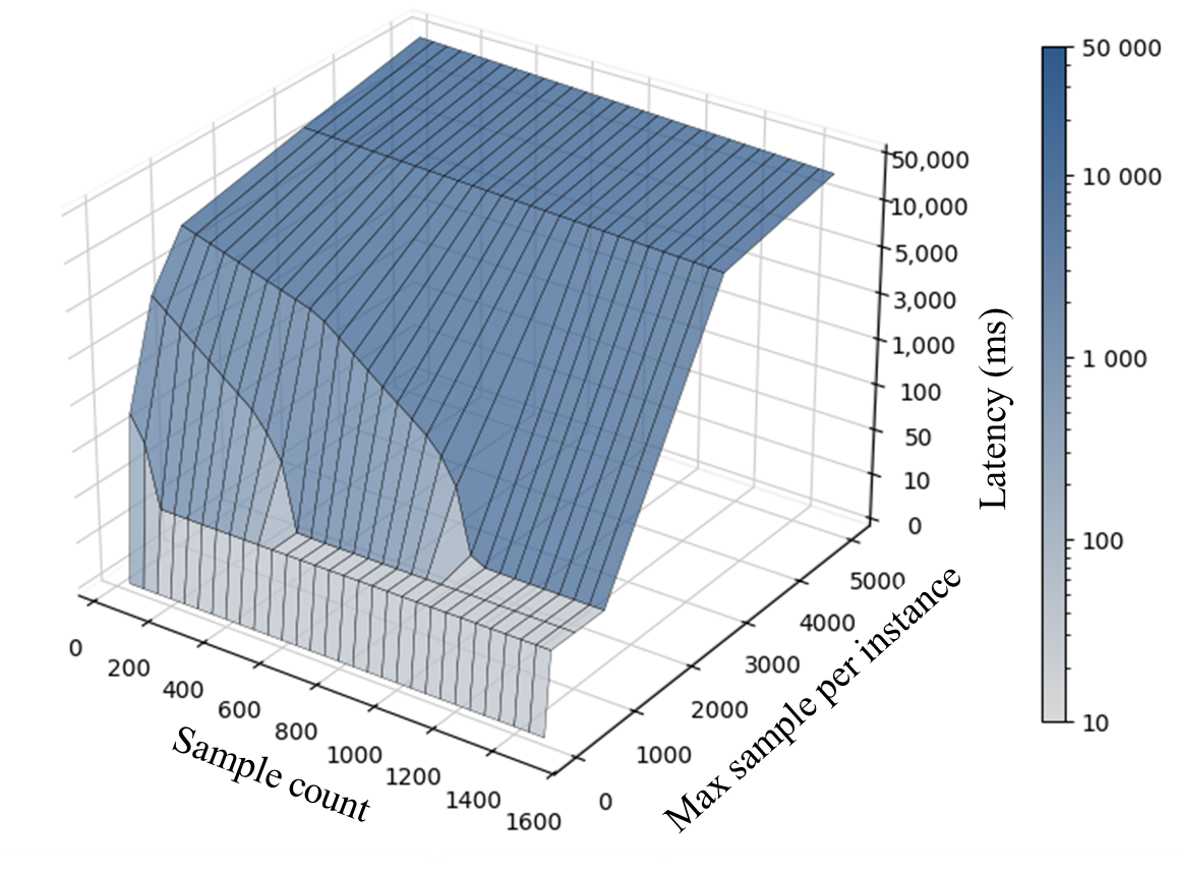}
\caption{Impact of HistoryCache Size on Latency of Late-join Subscriber}
\label{fig_2}
\end{figure}

Figure~\ref{fig_2} illustrates the initial latency observed when a Subscriber joins late, approximately 60~seconds after the Publisher, in a configuration where \qos{DURABL.kind=transient\_local}. 
The latency is plotted against varying values of HistoryCache size, as controlled by \qos{RESLIM.max\_samples\_per\_instance}. 
As shown in the graph, larger History Cache sizes result in an exponential increase in the number of previous samples transmitted during late join, which in turn causes a significant rise in network usage, thereby increasing the initial latency. 
In particular, when the HistoryCache size exceeds 4,000, the network becomes saturated and the system does not return to a low-latency state, as clearly observed in the experiment. 
Therefore, to ensure latency-critical late joining in a \qos{transient\_local} environment, the HistoryCache size must be carefully bounded according to the network capacity.

The \qos{DURABL} QoS policy has Incidental dependencies on \qos{PART}, \qos{ENTFAC}, and \qos{RDLIFE}. 
The delivery of previous samples by \qos{DURABL} to a late-joining reader occurs immediately after the Discovery phase matching. 
The \qos{PART} policy is modifiable at runtime, and any changes trigger re-execution of endpoint matching. 
If \qos{DURABL.kind$\geq$transient\_local}, newly matched pairs may be treated as late joiners, potentially causing redundant retransmission of previous samples. 
The \qos{ENTFAC} policy affects when a DataReader becomes active, which in turn determines the reference point from which \qos{DURABL} begins delivering stored samples.
If \qos{DURABL.kind=volatile}, initial sample availability is lost; therefore, \qos{transient\_local} or higher is generally recommended. 
Finally, if \qos{autopurge\_disposed\_samples\_delay} in \qos{RDLIFE} is set too short under \qos{DURABL.kind$\geq$transient}, late-arriving samples may be purged before the application has a chance to process them. 

The \qos{DEADLN} and \qos{LIVENS} QoS policies are responsible for monitoring real-time data exchange. 
The \qos{DEADLN} QoS policy exhibits a Critical self-dependency in SEDP RxO matching and has Conditional dependencies on \qos{RELIAB} and \qos{LIVENS}. 
It is recommended to use \qos{DEADLN} with \qos{RELIAB.kind=reliable}, since in \qos{best\_effort}, sample losses may go undetected, leading to false deadline-miss notifications. 
Moreover, setting \qos{LIVENS.lease\_duration} shorter than the \qos{DEADLN.period} may cause the deadline timer to stop upon liveliness expiration, preventing proper deadline monitoring.

\qos{DEADLN} has Incidental dependencies on \qos{PART} and \qos{DURABL}. 
If the \qos{PART.value} is modified at runtime, endpoint matching is re-executed and the deadline timer is reset, which may cause transient deadline-miss alarms for samples during the partition change. 
Furthermore, during retransmission of previous samples under \qos{DURABL}, repeated retransmissions can continuously reset the deadline timer, thereby masking actual real-time latency violations.

The \qos{LIVENS} QoS policy exhibits a Critical self-dependency in SEDP RxO matching and has a Conditional dependency on \qos{RELIAB}.
In practice, when using \qos{LIVENS.kind=manual\_by\_topic} for DataWriters with infrequent publications, it is recommended to configure \qos{RELIAB.kind=reliable}, since lost samples may cause the DataReader to falsely detect a loss of liveliness.
The \qos{LIVENS} policy also has an Incidental dependency on \qos{PART}. 
If the \qos{PART.value} is changed at runtime, the existing match between the DataWriter and DataReader is broken, even if the DataWriter continues asserting liveliness. 
Until rematching occurs, the DataReader no longer considers the DataWriter~alive. 

The \qos{HIST}, \qos{RESLIM}, and \qos{LFSPAN} QoS policies all directly govern the HistoryCache of DataWriters and DataReaders. 
Although their mutual influence is significant, their dependencies on other QoS policies are relatively limited. 
The combination of \qos{HIST} and \qos{RESLIM} constitutes a Critical Dependency. 
According to vendor guidelines, when \qos{HIST.kind=keep\_last}, the \qos{HIST.depth} must not exceed \qos{RESLIM.max\_samples\_per\_instance}. 
Conversely, in \qos{HIST.kind=keep\_all}, \qos{max\_samples\_per\_instance} defines the maximum number of samples that can be retained per instance. 

While \qos{HIST} and \qos{RESLIM} determine the number of samples that can be stored in the HistoryCache, \qos{LFSPAN} defines how long these samples can persist. 
\qos{LFSPAN} establishes a Conditional Dependency with \qos{HIST} and \qos{RESLIM}.
Regardless of \qos{HIST.depth}, samples are deleted once the \qos{LFSPAN.duration} expires. 
Conversely, if \qos{HIST.depth} is too small, even a long \qos{LFSPAN.duration} cannot ensure long-term retention of samples. 
In addition, within \qos{RESLIM}, the constraint \qos{max\_samples $\ge$ max\_samples\_per\_instance} must be satisfied. 
Violating this rule can lead to cache allocation errors or runtime exceptions, thereby constituting a Critical Dependency.

The \qos{OWNST} and \qos{DESTORD} QoS policies govern how a DataReader behaves when connected to multiple DataWriters. 
The \qos{OWNST} policy determines which DataWriter samples should be accepted. 
\qos{OWNST} exhibits a Critical self-dependency in SEDP RxO matching and a dependency on \qos{RELIAB}. 
If \qos{OWNST.kind=exclusive} is combined with \qos{RELIAB.kind=best\_effort}, packet loss can lead to false deadline-miss events, causing the DataReader to incorrectly switch ownership to another writer. 
In Fast DDS, using \qos{RELIAB.kind=best\_effort} with \qos{OWNST.kind=exclusive} may already be disallowed.
And \qos{OWNST} has Conditional dependencies on \qos{DEADLN} and \qos{LIVENS}. 
For \qos{OWNST.kind=exclusive}, ownership handover decisions are triggered by deadline-miss or liveliness-lost events. 
Therefore, if the \qos{DEADLN.period} or \qos{LIVENS.lease\_duration} is set too short, frequent and unintended ownership switches may occur; conversely, if they are set too long, timely ownership transitions may be delayed.

The \qos{DESTORD} QoS policy determines how a DataReader accepts samples from multiple DataWriters within its History Cache. 
First, \qos{DESTORD} exhibits a Critical self-dependency in SEDP RxO matching. 
It also has Conditional dependencies on \qos{HIST} and \qos{RESLIM}, which govern the structure and capacity of the HistoryCache. 
When configured as \qos{DESTORD.kind=by\_source\_timestamp}, the policy requires reordering samples by timestamp; however, if \qos{HIST.depth} or \qos{RESLIM.max\_samples\_per\_instance} is too small, there may be no space to insert earlier timestamped samples.
In particular, with the default setting \qos{HIST.depth=1}, only the most recent sample is retained, making it impossible for the intended ordering logic to function correctly.

The \qos{WDLIFE} and \qos{READER\_DATA\_LIFECYCLE} QoS policies control the removal of instances during the Disassociation phase. 
The \qos{WDLIFE} policy has a Conditional dependency on \qos{RELIAB}. 
This arises because dispose and unregister notifications are transmitted through RTPS DATA samples; when \qos{RELIAB.kind=best\_effort}, these samples may be lost, and the DataReader may not be informed of instance removal. 
In addition, \qos{WDLIFE} has an Incidental dependency on \qos{OWNST}. 
Under \qos{OWNST.kind=exclusive}, only the strongest DataWriter (the owner) is allowed to call dispose() for the deletion to be recognized. 
To avoid confusion in such cases, it is recommended to set \qos{WDLIFE.autodispose\_unregistered\_instances=false} and require the owner DataWriter to explicitly call dispose().

The operation of \qos{RDLIFE} requires either the detection that all associated DataWriters have been lost or the reception of a dispose sample. 
Accordingly, it has Conditional dependencies on both \qos{LIVENS} and \qos{WDLIFE}.
Once the \qos{LIVENS.\allowbreak lease\_duration} expires, the \qos{RDLIFE.autopurge\_no\_writer\_samples\_delay} timer begins, and after the configured period, the associated data are purged. 
To avoid premature deletion caused by false detection of liveliness loss or by failure to detect liveliness altogether, the \qos{LIVENS.lease\_duration} must be carefully configured. 
Moreover, if \qos{WDLIFE.\allowbreak autodispose\_unregistered\_instances=false} and the application does not explicitly call dispose(), the DataWriter will not issue a disposal notification. 
Consequently, the DataReader will not~transition~the~instance to \qos{not\_alive\_disposed}, rendering \qos{RDLIFE.autopurge\_disposed\_samples\_delay} inapplicable and ineffective for that instance.

\renewcommand{\arraystretch}{0.75}

\subsection{QoS Guard}
\begin{figure*}[!t]
\centering
\includegraphics[width=0.85\textwidth]{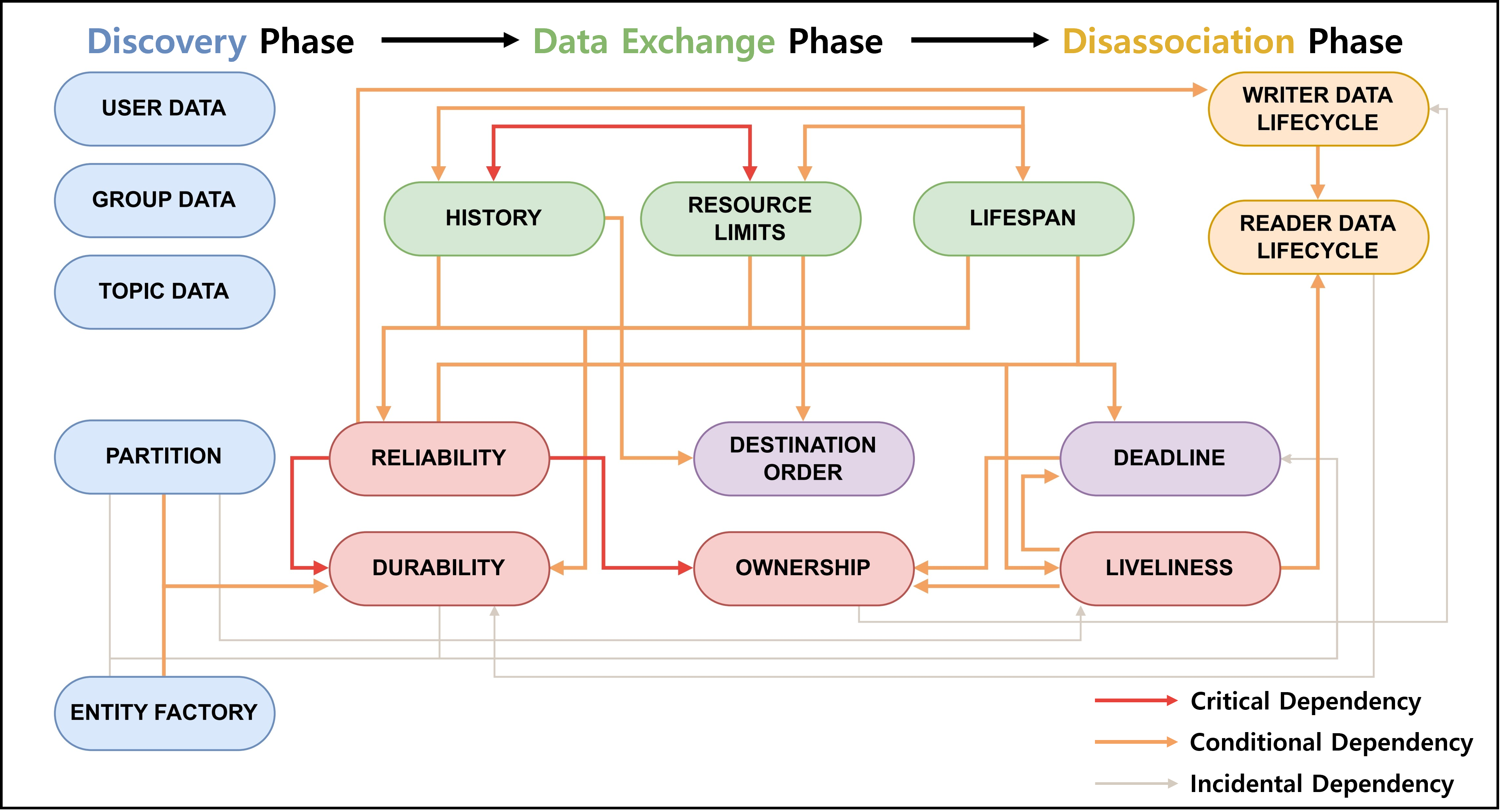}
\caption{QoS Policy Chain}
\label{fig_4}
\end{figure*}

Among the QoS policies defined in the OMG DDS standard, 16 are practically implemented by open-source DDS vendors. 
Based on the analysis of their interdependencies, we present the QoS Policy Chain in Figure~\ref{fig_4}. 
Each QoS policy is categorized by the lifecycle phase it affects, and Critical, Conditional, and Incidental dependencies are represented by red, orange, and gray lines, respectively. 
Building on the QoS Policy Chain, we propose QoS Guard, a tool for ROS~2 that automatically detects potential issues arising from policy interdependencies in Publisher and Subscriber XML profiles and suggests corrective configurations to the user. 
This section describes the validation algorithm implemented in QoS Guard. % and explains how the tool can be used.

QoS Guard employs a three-stage validation pipeline that spans the entire DDS communication lifecycle and respects the logical precedence of QoS policies. 
The first stage performs static intra-entity consistency checks: each DataWriter and DataReader is analyzed independently to identify inherently inconsistent configurations within the entity itself, regardless of the physical communication link.
Typical issues detected at this stage include intra-entity constraint violations and mismatches between QoS policies. 
Because these conflicts stem from static configuration parameters inside individual entities, they should be detected and corrected before entity~creation.

The second stage addresses RxO dependencies, where the QoS policies of the DataWriter and DataReader are compared for compatibility. 
By validating this compatibility during SEDP matching, QoS Guard helps prevent connection failures that would otherwise block communication initiation. 
The final stage evaluates dynamic inter-entity dependencies, focusing on the correctness of policy interactions that depend on runtime uncertainties such as variable latency and packet loss.
Although the impact of this stage is naturally limited in deterministic transport media such as Ethernet or shared memory, it becomes essential in wireless or mobile deployment scenarios. 
By progressing from static consistency checks to peer-level compatibility and finally to environment-specific robustness, QoS Guard enables ROS~2 users to detect configuration risks early and adapt DDS QoS profiles to the realities of their target deployment environments.

Table~\ref{table3} summarizes the 41 dependency-violation rules used by QoS Guard for validation. 
These rules are derived from the proposed QoS policy chain and comprise 19 static intra-entity dependencies for detecting internal conflicts, 8 RxO dependencies applicable to DataWriter-DataReader pairs, and 14 dynamic intra-entity dependencies triggered under runtime uncertainties such as latency and packet loss. 
The current DDS standard and major vendor implementations warn only about the 8 RxO dependencies and a few basic internal constraints; moreover, these RxO issues are difficult to detect without initiating the actual discovery process. 
In contrast, QoS Guard detects all conflicts without requiring a live DDS communication session. 
The tool applies the 41 rules to the provided QoS profiles and flags potential risks as errors or warnings, enabling users to resolve configuration issues before deployment.
QoS Guard is released as open-source software compatible with the ROS~2 Humble distribution, and installation instructions along with usage examples are available at \href{https://github.com/csi-dgist/qos_guard}{[GitHub]}.

\begin{table*}[htbp]
\centering
\caption{The 41 Dependency-Violation Rules Used in QoS Guard Validation}
\label{table3}
\renewcommand{\arraystretch}{1.25}
\begin{tabularx}{\textwidth}{c l p{8.5cm} l l c}
\toprule
\textbf{ID} & \textbf{Identifier} & \textbf{QoS Conflict Condition} & \textbf{Entity Scope} & \textbf{Dependency} & \textbf{Stage} \\
\midrule

1 & HIST $\leftrightarrow$ RESLIM
  & [HIST.kind = keep\_last] $\wedge$ \newline [HIST.depth $>$ RESLIM.max\_samples\_per\_instance]
  & - & Critical & 1 \\

2 & {RESLIM}~$\leftrightarrow$~{RESLIM} & 
    {RESLIM.max\_samples $<$ RESLIM.max\_samples\_per\_instance}
    & - & Critical & 1 \\

3 & {LFSPAN}~$\rightarrow$~{DEADLN} &
    LFSPAN.duration $<$ DEADLN.period
    & - & Critical & 1 \\

4 & {HIST}~$\rightarrow$~{DESTORD} & 
    [DESTORD.kind = by\_source\_timestamp] $\wedge$ \newline [HIST.kind = keep\_last] $\wedge$ [HIST.depth = 1]
    & DataReader & Conditional & 1 \\

5 & {RESLIM}~$\rightarrow$~{DESTORD} &
    [DESTORD.kind = by\_source\_timestamp] $\wedge$ \newline [HIST.kind = keep\_all] $\wedge$ \newline [RESLIM.max\_samples\_per\_instance = 1] 
    & DataReader & Conditional & 1 \\

6 & {HIST}~$\rightarrow$~{DURABL} &
    {[DURABL.kind $\ge$ transient\_local]} $\wedge$ [HIST.kind = keep\_last] $\wedge$ \newline
    {[HIST.depth $<$ (RTT / PP) + 2]} & DataWriter & Conditional & 1 \\

7 & {RESLIM}~$\rightarrow$~{DURABL} &
    [DURABILITY.kind $\ge$ transient\_local] $\wedge$  \newline [HISTORY.kind = keep\_all] $\wedge$ \newline
    [RESLIM.max\_samples\_per\_instance $<$ (RTT / PP) + 2]
    & DataWriter & Conditional & 1 \\

8 & {LFSPAN}~$\rightarrow$~{DURABL} &
    [DURABL.kind $\ge$ transient\_local] $\wedge$ [LFSPAN.duration $<$ RTT]
    & DataWriter & Conditional & 1 \\

9 & {HIST}~$\leftrightarrow$~{LFSPAN} &
    [HIST.kind = keep\_last] $\wedge$ \newline
    [LFSPAN.duration $>$ HIST.depth $\times$ PP]
    & DataWriter & Conditional & 1 \\

10 & {RESLIM}~$\leftrightarrow$~{LFSPAN} &
     [HIST.kind = keep\_all] $\wedge$ \newline
     [LFSPAN.duration $>$ RESLIM.max\_samples\_per\_instance $\times$ PP] 
     & DataWriter & Conditional & 1 \\

11 & {DEADLN}~$\rightarrow$~{OWNST} &
     [OWNST.kind = exclusive] $\wedge$ [DEADLN.period = $\infty$]
     & DataReader & Conditional & 1 \\

12 & {LIVENS}~$\rightarrow$~{OWNST} &
     [OWNST.kind = exclusive] $\wedge$ [LIVENS.lease\_duration = $\infty$] 
     & DataReader & Conditional & 1 \\

13 & {LIVENS}~$\rightarrow$~{RDLIFE} &
     [RDLIFE.autopurge\_nowriter\_samples\_delay $>$ 0] $\wedge$ \newline
     [LIVENS.lease\_duration = $\infty$] 
     & DataReader & Conditional & 1 \\

14 & {RDLIFE}~$\rightarrow$~{DURABL} &
     [DURABL.kind $\ge$ transient] $\wedge$ \newline
     [RDLIFE.autopurge\_disposed\_samples\_delay = 0] 
     & DataReader & Incidental & 1 \\

15 & {ENTFAC}~$\rightarrow$~{DURABL} &
     [DURABL.kind = volatile] $\wedge$ \newline
     [ENTFAC.autoenable\_created\_entities = false]
     & - & Incidental & 1 \\

16 & {PART}~$\rightarrow$~{DURABL} &
     [DURABL.kind $\ge$ transient\_local] $\wedge$ [PART.names $\ne$ \O] 
     & - & Incidental & 1 \\

17 & {PART}~$\rightarrow$~{DEADLN} &
     [DEADLN.period $>$ 0] $\wedge$ [PART.names $\ne$ \O]
     & - & Incidental & 1 \\

18 & {PART}~$\rightarrow$~{LIVENS} &
     [LIVENS.kind = manual\_by\_topic] $\wedge$ [PART.names $\ne$ \O] 
     & DataReader & Incidental & 1 \\

19 & {OWNST}~$\rightarrow$~{WDLIFE} &
     [WDLIFE.autodispose\_unregistered\_instances = true] $\wedge$ \newline
     [OWNST.kind = exclusive]
     & DataWriter & Incidental & 1 \\

% ==============================================

20 & {PART} ~$\leftrightarrow$~{PART} &
    [DataWriter.PART.names $\cap$ DataReader.PART.names] = \O
    & - & Critical & 2 \\

21 & {RELIAB}~$\leftrightarrow$~{RELIAB} & 
    DataWriter.RELIAB.kind $<$ DataReader.RELIAB.kind
    & - & Critical & 2 \\

22 & {DURABL}~$\leftrightarrow$~{DURABL} &
    DataWriter.DURABL.kind $<$ DataReader.DURABL.kind
    & - & Critical & 2 \\

23 & {DEADLN}~$\leftrightarrow$~{DEADLN} &
    DataWriter.DEADLN.period $>$ DataReader.DEADLN.period
    & - & Critical & 2 \\

24 & {LIVENS}~$\leftrightarrow$~{LIVENS} &
    [DataWriter.LIVENS.kind $<$ DataReader.LIVENS.kind] $\vee$ \newline
    [DataWriter.LIVENS.lease\_duration $>$ DataReader.LIVENS.lease\_duration] 
    & - & Critical & 2 \\

25 & {OWNST}~$\leftrightarrow$~{OWNST} &
    DataWriter.OWNST.kind $\ne$ DataReader.OWNST.kind
    & - & Critical & 2 \\

26 & {DESTORD}~$\leftrightarrow$~{DESTORD} &
    [DataWriter.DESTORD.kind $<$ DataReader.DESTORD.kind] 
    & - & Critical & 2 \\

27 & {WDLIFE}~$\rightarrow$~{RDLIFE} &
    [WDLIFE.autodispose\_unregistered\_instances = false] $\wedge$ \newline
    [RDLIFE.autopurge\_disposed\_samples\_delay $>$ 0] 
    & - & Conditional & 2 \\

28 & {RELIAB}~$\rightarrow$~{DURABL} &
    [DURABL.kind $\ge$ transient\_local] $\wedge$ [RELIAB.kind = best\_effort] 
    & - & Critical & 3 \\

29 & {HIST}~$\rightarrow$~{RELIAB} &
    [RELIAB.kind = reliable] $\wedge$ [HIST.kind = keep\_last] $\wedge$ \newline
    [HIST.depth $<$ (RTT/PP) + 2] 
    & DataWriter & Conditional & 3 \\

30 & {RESLIM}~$\rightarrow$~{RELIAB} &
    [RELIAB.kind = reliable] $\wedge$ [HIST.kind = keep\_all] $\wedge$ \newline
    [RESLIM.max\_samples\_per\_instance $<$ (RTT/PP) + 2] &
    DataWriter & Conditional & 3 \\

31 & {LFSPAN}~$\rightarrow$~{RELIAB} &
    [RELIAB.kind = reliable] $\wedge$ [LFSPAN.duration $<$ RTT] 
    & DataWriter & Conditional & 3 \\

32 & {RELIAB}~$\rightarrow$~{OWNST} &
    [OWNST.kind = exclusive] $\wedge$ [RELIAB.kind = best\_effort] 
    & - & Conditional & 3 \\

33 & {RELIAB}~$\rightarrow$~{DEADLN} &
    [DEADLN.period $>$ 0] $\wedge$ [RELIAB.kind = best\_effort] 
    & - & Conditional & 3 \\

34 & {LIVENS}~$\rightarrow$~{DEADLN} &
    [DEADLN.period $>$ 0] $\wedge$ [LIVENS.lease\_duration $<$ DEADLN.period] 
    & DataReader & Conditional & 3 \\

35 & {RELIAB}~$\rightarrow$~{LIVENS} &
    [LIVENS.kind = manual\_by\_topic] $\wedge$ [RELIAB.kind = best\_effort] 
    & - & Conditional & 3 \\

36 & {DEADLN}~$\rightarrow$~{OWNST} &
    [OWNST.kind = exclusive] $\wedge$ [DEADLN.period $<$ 2 $\times$ PP]
    & DataReader & Conditional & 3 \\

37 & {LIVENS}~$\rightarrow$~{OWNST} &
    [OWNST.kind = exclusive] $\wedge$ [LIVENS.lease\_duration $<$ 2 $\times$ PP] 
    & DataReader & Conditional & 3 \\

38 & {RELIAB}~$\rightarrow$~{WDLIFE} &
    [WDLIFE.autodispose\_unregistered\_instances = true] $\wedge$ \newline 
    [RELIAB.kind = best\_effort] 
    & DataWriter & Conditional & 3 \\

39 & {HIST}~$\rightarrow$~{DURABL} &
    [DURABL.kind $\ge$ transient\_local] $\wedge$ [HIST.kind = keep\_last] $\wedge$ \newline
    [HIST.depth $>$ (RTT/PP) + 2] 
    & DataWriter & Incidental & 3 \\

40 & {RESLIM}~$\rightarrow$~{DURABL} &
    [DURABL.kind $\ge$ transient\_local] $\wedge$ {[HIST.kind = keep\_all]} $\wedge$ \newline
    [RESLIM.max\_samples\_per\_instance $>$ (RTT/PP) + 2] 
    & DataWriter & Incidental & 3 \\

41 & {DURABL}~$\rightarrow$~{DEADLN} &
    [DEADLN.period $>$ 0] $\wedge$ [DURABL.kind $\ge$ transient\_local] 
    & - & Incidental & 3 \\

\bottomrule
\end{tabularx}
\end{table*}

\newpage
\section{Conclusion}
\label{sec05}
ROS~2 is the de facto standard for robot development, and users simply want it to just work. 
However, the QoS policies that ROS~2 exposes are not especially user-friendly, increasing the likelihood of trial-and-error. 
To bridge this gap, this paper proposes a lifecycle-based QoS tutorial that explains how 16 key QoS policies operate across discovery, data exchange, and disassociation, illustrated with mobile-robot examples.
It also introduces a Qos policy chain that systematizes inter-policy relationships into three tiers, making them easy to grasp.
Finally, it presents QoS Guard, a static validation tool that encodes 41 dependency rules to scan DDS XML profiles offline, flag conflicts before deployment, and guide safe configuration.
Together, these components replace ad hoc tuning with structured guidance and pre-deployment checks, reducing runtime risks such as data loss and network saturation and improving the reliability and deployment efficiency of ROS~2-based distributed systems.

As a limitation, QoS Guard currently suggests only suitable parameter ranges for continuous QoS values (e.g., \qos{DEADLINE.period}, \qos{LIFESPAN.duration}, \qos{HISTORY.depth}) and does not automatically compute quantitatively optimal values based on network conditions, task frequency, or packet loss rate. 
Because these values depend on external variables such as link bandwidth, latency, and packet loss, a more sophisticated tuning mechanism is required. 
In future work, we will extend QoS Guard into a QoS Optimizer that accepts user-defined constraints-such as target latency, availability, and memory limits-and dynamically recommends optimal QoS configurations by incorporating real-time network characteristics.

\newpage
\bibliographystyle{IEEEtran}
\bibliography{main}

% \newpage

 

% \end{IEEEbiography}



\end{document}